\documentclass[aps,pra,nobibnotes,twocolumn,superscriptaddress]{revtex4-1}
\usepackage{graphicx}
\usepackage{mathrsfs}
\usepackage{bm}
\usepackage{amsmath}
\usepackage{dcolumn}
\usepackage{epstopdf}
\usepackage{dsfont}
\usepackage{amssymb}
\usepackage{tabularx}
\usepackage{array}
\usepackage{float}
\usepackage{color}
\usepackage{epstopdf}
\usepackage{mathrsfs}
\usepackage[colorlinks, linkcolor=blue,anchorcolor=blue,citecolor=blue,urlcolor=blue]{hyperref}

\begin{document}

\preprint{APS/123-QED}

\title{Quantum phase transitions from competing short- and long-range interactions on a $\pi$-flux lattice}

\author{Xingchuan Zhu}
\affiliation{Interdisciplinary Center for Fundamental and Frontier Sciences, Nanjing University of Science and Technology, Jiangyin, Jiangsu 214443, P. R. China}

\author{Yiqun Huang}
\affiliation{Department of Physics, Beijing Normal University, Beijing, 100875, China}

\author{Huaiming Guo}
\email{hmguo@buaa.edu.cn}
\affiliation{School of Physics, Beihang University, Beijing, 100191, China}

\author{Shiping Feng}
\affiliation{Department of Physics, Beijing Normal University, Beijing, 100875, China}

\pacs{ 03.65.Vf, 
 67.85.Hj 
 73.21.Cd 
 }

\begin{abstract}
Quantum phase transitions from the cluster-charge interaction, which is composed of competing short- and long-range interactions, are investigated on a $\pi$-flux lattice by using the mean-field theory and determinant quantum Monte Carlo (DQMC) simulations. Both methods identify a plaquette-dimer phase, which develops from a finite interaction strength. While its signature in DQMC is relatively weak, a obvious antiferromagnetic transition is revealed in the spin structure factor instead. The corresponding critical interaction and exponents are readily obtained by finite-size scalings, with the plaquette-dimer structure factor that can also be well scaled. These results suggest a possible deconfined quantum critical point between the plaquette-dimer and antiferromagnetic phases driven by the cluster-charge interaction on a $\pi$-flux lattice.

\end{abstract}

\maketitle

\section{Introduction}

Dirac semimetal (SM) in two dimension (2D) has attracted intense interests in condensed matter physics~\cite{graphene2009,Geim2009,Young2015,Cayssol2013,Wang2017115138,Young2017,Wehling2014,Vafek2014}. The low-energy electronic states of this class of quantum matters can be effectively described by Dirac equation. The resulting linear dispersion relation leads to many exotic physical phenomena. Besides the well-known graphene, various 2D Dirac materials have been predicted and discovered experimentally up to now~\cite{Zhou2016,Cai2015,Wang2017,Khusnutdinov2018,Bykov2021,Zhang2021,Yang2017,Silicon2009,Silicene2011}.

The emergence of massless Dirac fermions is usually protected by specific lattice symmetries. When the lattice is distorted by various periodic perturbations, the Dirac semimetal can spawn interesting insulating phases. A recent key theoretical advance is to gap the Dirac semimetal with spin-orbit coupling, which has led to the discovery of time-reversal-invariant topological insulators~\cite{Kane2005,Kane2010,SCZhang2011}. Other insulating phases can be generated by introducing various kinds of dimerizations of the hopping amplitudes and charge-density-wave modulations of the on-site energies~\cite{Semenoff,Mudry2007,Guo2009}. The above periodic orders are usually driven by interactions through spontaneous symmetry breaking. When the repulsive interactions of different ranges coexist, a rich phase diagram, composed of various charge, spin and topological ordered phases, will be obtained\cite{Raghu2008,Weeks2010,Fiete2010,Scherer2015}. The abundant phase transitions therein make the system to become an ideal platform to investigate the exotic quantum criticality.

While most quantum critical behaviors can be well described by Landau's theory of phase transitions~\cite{Hohenberg2015,Wilson1974}, recent theoretical studies have led to the notion that Landau's description is insufficient. Particularly, a new class of phase transition termed as deconfined quantum critical point (DQCP) is established based on 2D antiferromagnet~\cite{Senthil2004,Senthil20041490,Wang2017031051,Sandvik2007,Huang2019,Levin2004,Nahum2015}. DQPC occurs in a phase transition between the N\'{e}el and valence bond solid (VBS) phases. Although either a first-order phase transition or phase coexistence is expected according to Landau's theory, the actual transition is a direct second-order one, thus should be described by new theory, which has been termed as DQCP.

Deconfined quantum criticality may also exist in fermionic systems~\cite{Li2017,Liu2022,Xu2019,wang2021,Sato2017,Li2019}. While the antiferromagnetic (AF) order can be induced by the on-site Hubbard repulsion, the VBS phase can be stabilized by competing short- and long-range interactions. Indeed convincing evidences of DQCP between the AF and VBS phases have been demonstrated in the extended Hubbard model of fermions on
honeycomb lattice~\cite{Sato2017,Li2019}. The studies on DQCP in fermionic systems are much difficult since the quantum Monte Carlo (QMC) simulations are usually restricted to smaller lattice sizes than those of spin Hamiltonians. Therefore, additional specially-designed interactions need to be included to stabilize the ordered phases (especially the VBS phase) in a larger parameter regime~\cite{Li2019}.

In the initial stage of modeling the strong correlation physics in twisted bilayer graphene, the cluster-charge interaction, which has simple form and is feasible for sign-problem-free DQMC simulations, has been proposed as an effective interaction on honeycomb lattice~\cite{Xu2018,Da2019}. Interestingly it is found the Kekul\'{e} VBS is stabilized in a wide range of interaction strength. Although there exists a transition from the VBS to AF phases, it is shown to be a first-order one. As the other realization of 2D Dirac semimetal, the $\pi$-flux lattice has a distinct lattice symmetry and coordination number from the honeycomb lattice~\cite{Li2015,Guo2018235152,Otsuka2016,Otsuka2002,Chang2012,assaad2015,Rosenberg2009,Jia2013,Guo2018,Ouyang2021}. Thus it is natural to ask what kind of VBS may be stabilized, and whether a DQCP may be realized by the cluster-charge interaction on the $\pi$-flux lattice.

In this paper, the quantum phase transitions from competing short- and long-range interactions constituting the cluster-charge interaction are investigated on a $\pi$-flux lattice. The mean-field theory predicts the appearance of plaquette-dimer phase from a finite interaction strength. While the subsequent DQMC simulations find signatures of the above SM-VBS transition, the lattice sizes accessed are not large enough to characterize the VBS transitions due to the fragility of the plaquette-dimer phase. A clear AF transition is revealed instead, whose critical interaction and exponents are estimated by finite-size scalings. Furthermore, we find the structure factor of the VBS phase can be scaled satisfactorily with the above critical values. Our results suggest a possible DQCP between the plaquette-dimer and AF phases driven by the cluster-charge interaction on a $\pi$-flux lattice.

This paper is organized as follows. Section II introduces the model we will investigate, along with our computational methodology. Section III presents the results
from the mean-field theory. Section IV uses DQMC simulations to study the quantum phase transitions of the interacting Hamiltonian. Section V includes the conclusions and discussions.

\section{The model and method}

We start with the following Hamiltonian describing interacting spin-$1/2$ fermions on a $\pi$-flux lattice~\cite{Rosenberg2009,Jia2013}:
\begin{eqnarray}\label{Hmodel}
H=-\sum_{\langle ij\rangle,\sigma}(t_{ij}c^{\dagger}_{i\sigma}c_{j\sigma}+\textrm{H.c.} )+U\sum_{\square}(Q_{\square}-2)^2,
\end{eqnarray}
where the sum on $\sigma$ runs over spins $\sigma=\uparrow,\downarrow$, and $\langle ij\rangle$ denotes the nearest-neighbor (NN) pairs; $c^{\dagger}_{i\sigma}$ and $c_{i\sigma}$ are creation and annihilation operators of electrons with spin $\sigma$ on a given site $i$; the cluster charge $Q_{\square}=\sum_{i\in\square}\frac{n_i}{2}$ is defined as the total charge on the four sites of each plaquette with $n_i=\sum_{\sigma}c^{\dagger}_{i\sigma}c_{i\sigma}$ the total number operator of electrons on each site; $U$ is the interaction strength. To avoid the cluster-charge interaction, four electrons per plaquette are energetically favored, when the system is exactly at half filling.

The first term in Eq.(1) is noninteracting, describing the electrons on a square lattice subjected to a magnetic field. The magnetic flux per plaquette is one half of a magnetic flux quantum $\Phi_{0} = hc/e$. As a consequence, when an electron hopping along the four bonds constituting a plaquette in one direction, a total phase $\pi$ is picked up when it returns to the starting point. We choose the Landau gauge so that all hopping amplitudes in the $x$ direction are $t_x=t$, while the hopping signs along the $y$-direction are staggered, i.e., $t_y = (-1)^{i_x}t= \pm t$ ($i_x$ represents the $x$-coordinate of the $i$-th site). The resulting lattice is composed of two sublattices, and the unit cell contains two sites $A$ and $B$. In the reciprocal
space, within the reduced Brillouin zone $(|k_x| \leq \pi/2, |k_y| \leq
\pi)$, the noninteracting Hamiltonian can be written as
\begin{align}
H_0 &=\sum_{\bf{k}\sigma}
\psi^{\dagger}_{\bf{k}\sigma}
{\cal H}_0({\bf k})
\psi_{\bf{k}\sigma},
\end{align}
with the basis $\psi_{\bf{k}\sigma} =\left(c^{\phantom{\dagger}}_{A\sigma}, c^{\phantom{\dagger}}_{B\sigma}\right)^{T}$
and the Hamiltonian in the momentum space
\begin{align}
{\cal H}_0({\bf k}) =
\left( \begin{array}{cc}
 -2t \, {\rm cos} k_y & +2t \, {\rm cos}k_x \\
 +2t \, {\rm cos} k_x & +2t \, {\rm cos}k_y \\
\end{array} \right).
\end{align}
The energy spectrum is given by
\begin{align}
E_{\bf k} = \pm
\sqrt{4t^{2}(\cos^2 k_x+\cos^2 k_y)},
\end{align}
which is symmetric around the Fermi level, and gapless at the two inequivalent Dirac points located at ${\bf K}_{\pm}=(\pi/2,\pm \pi/2)$.

The interacting term in the Hamiltonian Eq.(1) contains various kinds of short- and long-range interactions, which is more easily seen through an  expanding of the cluster-charge interaction~\cite{Xu2018},
\begin{align}\label{clusterin}
&U\sum_{\square}(Q_{\square}-2)^2= 2U\sum_{i}n_{i\uparrow}n_{i\downarrow}+U\sum_{\langle ij\rangle}n_{i}n_{j}  \\
&+\frac{1}{2}U\sum_{\langle\langle ij\rangle\rangle}n_{i}n_{j}-7U\sum_{i}n_{i}+4UN_{s}, \nonumber
\end{align}
with $N_s$ the total number of sites on the lattice. $\langle\langle ij\rangle\rangle$ means the next-nearest-neighbor (NNN) interactions.
Thus the system includes on-site, NN, and NNN repulsions, and  the interaction strength ratio from on-site to NNN ones is $4:2:1$.

\begin{figure}[htbp]
\centering \includegraphics[width=7.5cm]{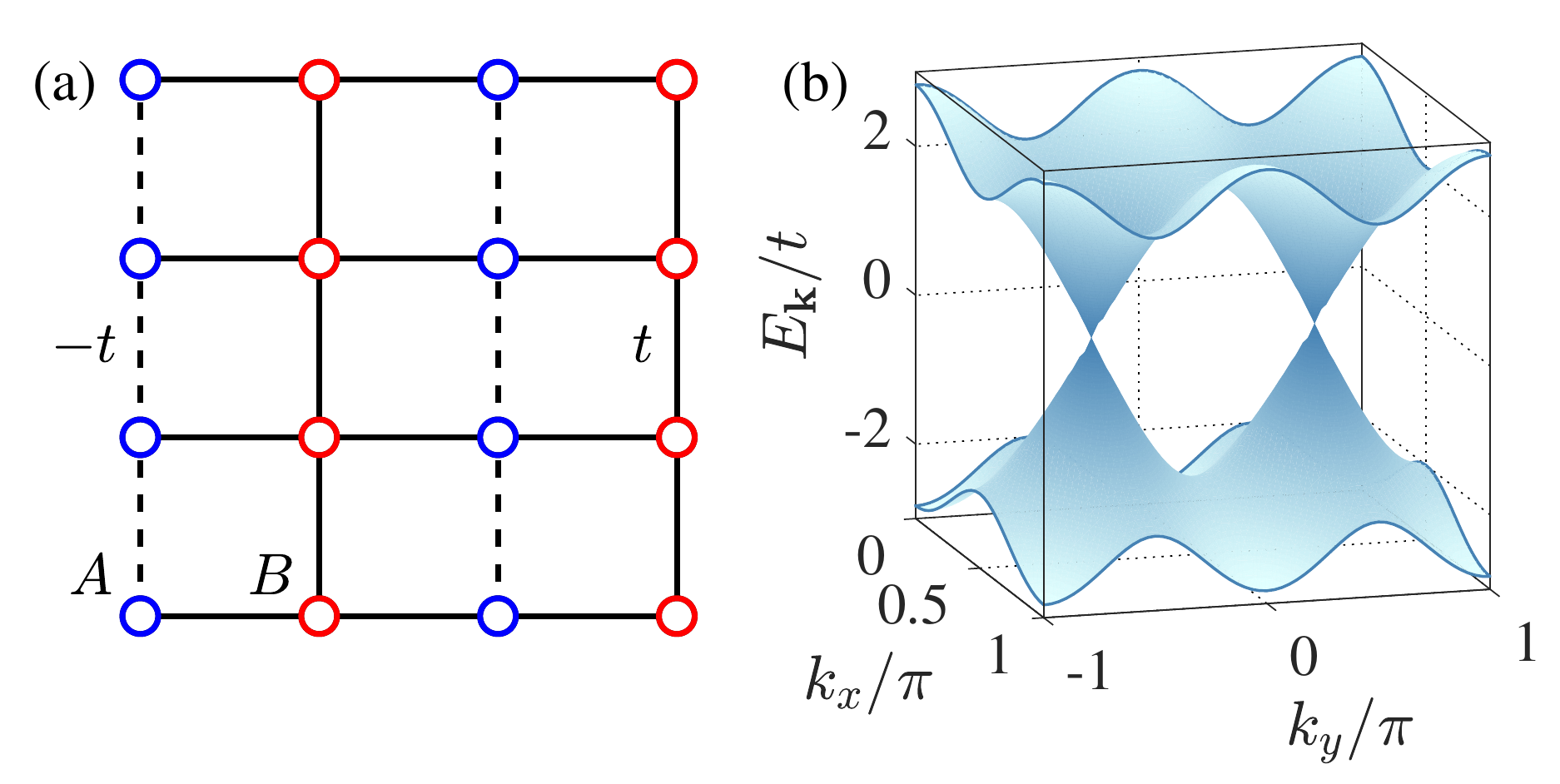} \caption{(a) The $\pi$-flux model on the square lattice. The unit cell is composed of two inequivalent sites $A$ and $B$. The NN hopping amplitudes may be $t$ (solid line) or $-t$ (dashed line), depending on the directions and positions of the bonds. (b) Band structure of the $\pi$-flux square lattice with two Dirac cones at the momentum points $(k_x,k_y)=(\pi/2,\pm \pi/2)$. }
\label{fig1}
\end{figure}

At finite strength of $U$, Eq.(1) is solved numerically via DQMC, where one decouples the two-body interaction in perfect square form through the introduction of an auxiliary Hubbard-Stratonovich field, which is integrated out stochastically~\cite{Blankenbecler1981,White1989,White1989839}. The only errors are those associated with the statistical sampling, the finite spatial lattice size, and the inverse temperature discretization. These errors are well controlled in the sense that they can be systematically reduced as needed, and further eliminated by appropriate extrapolations. At half filling, the simulaitons are free of sign problems due to the presence of particle-hole symmetry~\cite{Loh1990,Troyer2005,Iglovikov2015,Li2015241117}. Thus we can access low enough temperatures, necessary to determine the ground-state properties on finite-size lattices.
In the following, we use the inverse temperature discretization $\Delta\tau=0.1$, and the simulations are carried out on $L \times L$ lattices with the linear size $L$ up to $24$.

\section{The mean-field approximation}
We first treat the Hamiltonian in Eq.(1) using the mean-field approximation, which should be helpful to identify the possible ordered quantum phases. Here we can deal with various values of the interaction strengths, thus start from the following general interaction terms,
\begin{align}\label{clusterin}
H_{int}= U_0\sum_{i}n_{i\uparrow}n_{i\downarrow}+V_1\sum_{\langle ij\rangle}n_{i}n_{j}+V_2\sum_{\langle\langle ij\rangle\rangle}n_{i}n_{j},
\end{align}
where $U_0, V_1, V_2$ are the strengths of the on-site, NN and NNN interactions, respectively.
All interactions are decoupled in the on-site channel as~\cite{Weeks2010,Fiete2010},
\begin{align}
n_{i,\uparrow}n_{i,\downarrow}&=\langle n_{i,\uparrow} \rangle n_{i,\downarrow} + \langle n_{i,\downarrow} \rangle n_{i,\uparrow}-\langle n_{i,\uparrow}\rangle \langle n_{i,\downarrow}\rangle, \\ \nonumber
n_i n_j&=\langle n_i\rangle n_j + \langle n_j\rangle n_i - \langle n_i\rangle \langle n_j\rangle.
\end{align}

\begin{figure}[htbp]
\centering \includegraphics[width=7.cm]{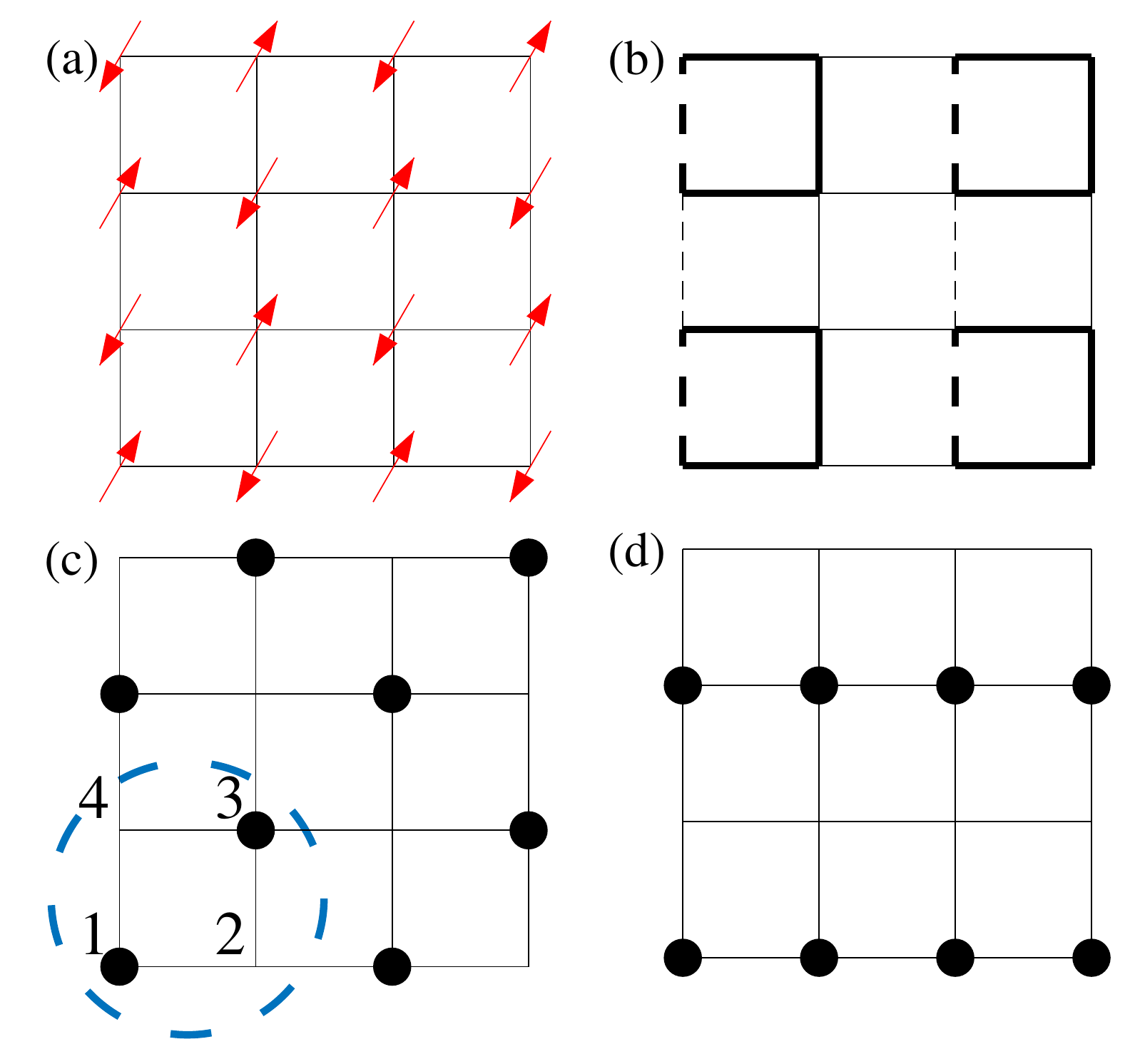} \caption{Schematics of the possible ordered phases on square lattice: (a) AF insulator, (b) plaquette dimerization, (c) staggered CDW, and (d) striped CDW.  In the presence of all four orders, the unit cell is enlarged to contain four sites, each of which is marked with an integer index [see (c)]. The configurations in (a), (c), and (d) are solely favored by the on-site, NN, and NNN interactions, respectively.}
\label{geometry}
\end{figure}

The on-site Hubbard, NN and NNN repulsive interactions favor antiferromagnetism, staggered and striped charge-density-waves (CDWs), respectively. We introduce three order parameters $\sigma, \rho, \nu$ to characterize the above three phases.
Meanwhile, the unit cell is enlarged to have four sites in the presence of the above charge and spin configurations. Then we have the following ansatz for the average density $\rho_i=\langle n_i \rangle$ on each site,
\begin{eqnarray}\label{rhoi}
\rho_i=1+(-1)^{i-1}\rho+(-1)^{m_i}\nu, \\
\rho_{i,\uparrow}=\rho_i/2+(-1)^{i-1}\sigma, \nonumber\\
\rho_{i,\downarrow}=\rho_i/2-(-1)^{i-1}\sigma,  \nonumber
\end{eqnarray}
where $i=1,2,3,4$ labels the sites in a unit cell, and $m_1=m_2=0$, $m_3=m_4=1$.

To incorporate possible valence-bond ordered phases, we also consider a bond decoupling
channel for the NN interaction,
\begin{align}
&n_i n_j  =c^{\dagger}_ic_ic^{\dagger}_j c_j \nonumber \\&
=-\langle c^{\dagger}_ic_j\rangle c^{\dagger}_j c_i -\langle c^{\dagger}_j c_i\rangle c^{\dagger}_ic_j +\langle c^{\dagger}_j c_i\rangle \langle c^{\dagger}_ic_j\rangle.
\end{align}
While there exist various kinds of dimer patterns on square lattice, here we focus on the spontaneous plaquette dimerization [see Fig. \ref{geometry}(b)], which we will demonstrate to emerge out of the competing interactions in Eq.(6) thereafter.

In the momentum space, the mean-field Hamiltonian is
\begin{align}
H_{M F}=\sum_{\mathbf{k}\sigma} \psi_{\mathbf{k}}^{\dagger} [{\mathcal{H} }^{1}_{\sigma}(\mathbf{k})+{\mathcal{H} }^{2}_{\sigma}(\mathbf{k}) ] \psi_{\mathbf{k}}+E_{0}.
\end{align}
Here $\psi_{\mathbf{k}}=\left(c_{1, \mathbf{k}}, c_{2, \mathbf{k}}, c_{3, \mathbf{k}},c_{4, \mathbf{k}}\right)^{T} $ is a four-element basis.
${\mathcal{H} }^{1}_{\sigma}(\mathbf{k})$ is spin-dependent, and for the up-spin subsystem (the formula for the down-spin copy is similar), it writes as,
\begin{align}\label{eq1}
{\mathcal{H} }^{1}_{\uparrow}(\mathbf{k})=\left[
  \begin{array}{cccc}
   h_{11} & t(k_x)^* & 0 & -t(k_y)^* \\
    t(k_x) &  h_{22} & t(k_y)^* & 0 \\
    0 & t(k_y) & h_{33} & t(k_x) \\
    -t(k_y) & 0 & t(k_x)^* &  h_{44} \\
  \end{array}
\right],
\end{align}
with $t(k_{\alpha})=t(1+e^{ik_{\alpha}})$ ($\alpha=x,y$) and
\begin{align}\label{eq1}
h^{\uparrow}_{ii}= & (-1)^{i-1}\left( \frac{U_0}{2}-4V_1+4V_2 \right)\rho \\ \nonumber
&+(-1)^{m_i}\left(\frac{U_0}{2}-4V_2\right)\nu \\ \nonumber
& +(-1)^i U_0\sigma+\left(\frac{U_0}{2}+4V_1+4V_2\right) \nonumber.
\end{align}
${\mathcal{H} }^{2}_{\sigma}(\mathbf{k})$ is decoupled from the NN interaction, and does not depend on the spin index,
\begin{align}\label{eq1}
&{\mathcal{H} }^{2}_{\sigma}(\mathbf{k})=
&\left(
  \begin{array}{cccc}
   0 & h_{\chi}(k_x)^* & 0 & -h_{\chi}(k_y)^* \\
    h_{\chi}(k_x) &  0 & h_{\chi}(k_y)^* & 0 \\
    0 & h_{\chi}(k_y) & 0 & h_{\chi}(k_x) \\
    -h_{\chi}(k_y) & 0 & h_{\chi}(k_x)^* &  0 \\
  \end{array}
\right),
\end{align}
where $\chi_1, \chi_2=-\langle c_{i\sigma}^{\dagger}c_{j\sigma} \rangle$ are for thick and thin bonds in Fig. \ref{geometry}(b); $h_{\chi}(k_{\alpha})=V_1(\chi_1+\chi_2e^{ik_{\alpha}})$ with $\alpha=x,y$.
Here the constant is,
\begin{align}\label{eq1}
E_0 &=(8V_2 - U_0)\nu^2 + (8V_1 - U_0 - 8V_2)\rho^2 \nonumber\\ &+ 4U_0\sigma^2 -
U_0 - 8V_1 - 8V_2+4V_1[\chi_1^2+\chi_2^2].
\end{align}

We can diagonalize the total Hamiltonian and obtain the dispersion with four branches
\begin{align}\label{eq1}
E_i({\bf k}) =&D+(-1)^{i}\frac{1}{\sqrt{2}}\sqrt{A+(-1)^{m_{i}+1}B+C},\nonumber \\
A =&a^2+b^2+4(t_{1}^2+t_{2}^2), \nonumber \\
B =&(a+b)\sqrt{(a-b)^2+4(t_{1}^2+t_{2}^2)+8t_{1}t_{2}cos(k_{x})}, \nonumber \\
C =&4t_{1}t_{2}[cos(k_{x})+cos(k_{y})], \nonumber \\
D =&\frac{U_0}{2}+4V_{1}+4V_{2},
\end{align}
with
\begin{align}\label{eq1}
a =&\left(  \frac{U_0}{2}-4V_{1}+4V_{2} \right)\rho+\left(\frac{U_0}{2}-4V_2\right)\nu-U_0\sigma, \nonumber \\
b =&-\left(  \frac{U_0}{2}-4V_{1}+4V_{2} \right)\rho+\left(\frac{U_0}{2}-4V_2\right)\nu+U_0\sigma.
\end{align}

Minimizing the total energy $E_{tot}=\sum_{i,{\bf k}}E_i({\bf k})+E_0$, the order parameters of the ground state satisfy the following self-consistent equations:
\begin{align}\label{eq1}
\rho=&-\frac{1}{2(8V_1-8V_2-U_0)}\frac{\partial E_{tot}}{\partial \rho}, \\
\nu=&-\frac{1}{2(8V_2-U_0)}\frac{\partial E_{tot}}{\partial \nu}, \nonumber\\
\sigma=&-\frac{1}{8U_0}\frac{\partial E_{tot}}{\partial \sigma}, \nonumber\\
\chi_1=&-\frac{1}{8V_1}\frac{\partial E_{tot}}{\partial \chi_1}, \nonumber\\
\chi_2=&-\frac{1}{8V_1}\frac{\partial E_{tot}}{\partial \chi_2}. \nonumber
\end{align}

\begin{figure}[htbp]
\centering \includegraphics[width=8.cm ] {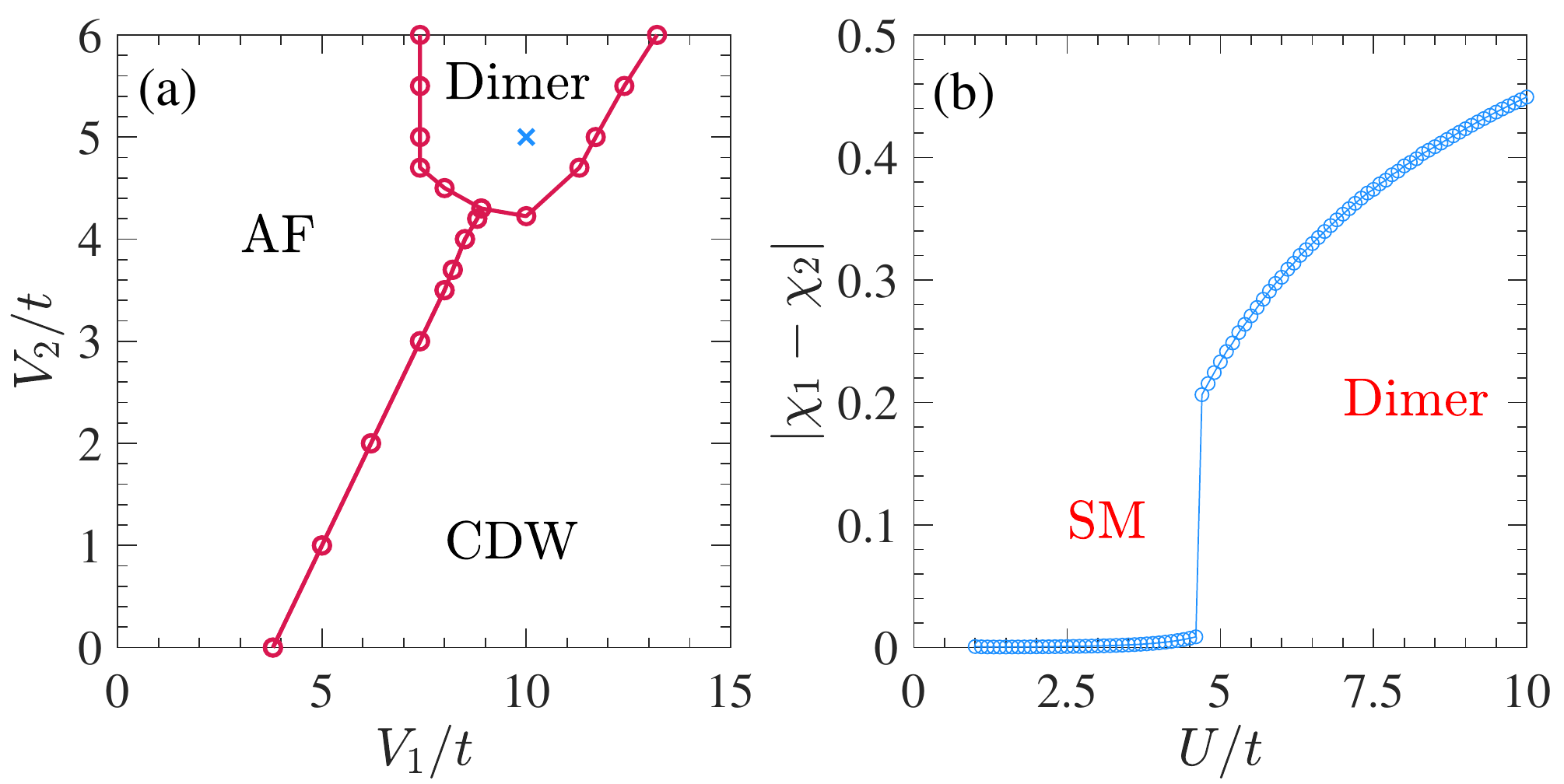} \caption{(a) The phase diagram in the $(V_1, V_2)$ plane at fixed $U_0/t=20$. The blue cross marks the point when the interactions can be written in perfect square form of the cluster charge. (b) The order parameter $|\chi_1-\chi_2|$, which characterizes the plaquette-dimer phase, as a function of the cluster-charge interaction $U$.}
\label{pifluxorderphase}
\end{figure}

The order parameters can be obtained by numerically solving the above equations. In Fig. \ref{pifluxorderphase}, we plot the phase diagram in the $(V_1, V_2)$ plane at fixed $U_0/t=20$. Three kinds of phases, including: AF, CDW and plaquette-dimer states, are revealed. In the absence of $V_1$ and $V_2$, the system is an AF insulator at $U_0/t=20$. A phase transition from AF to CDW is driven by the NN interaction, and the critical interaction increases with $V_2$. When $V_2$ is large enough, there appears a region of plaquette-dimer phase around $V_1=2V_2$. Particularly, at $V_1=2V_2=U_0/2$ when the interactions can be written in a perfect square form of the cluster charge, the system is in the plaquette-dimer ordered state. We then investigate the quantum phase transition driven by the cluster-charge interaction $U$ (the ratio of the interaction strengths to be $U_0:V_1:V_2=4:2:1$, and $U_0=2U$). As $U$ increases, all other order parameters remain vanished except for $|\chi_1-\chi_2|$ characterizing the plaquette dimerization. $|\chi_1-\chi_2|$ jumps to a finite and large value at $U/t=4.6$, marking the occurrence of a phase transition from SM to plaquette dimer.

\section{Results from the DQMC simulations}
With the mean-field insights into the ground-state properties of the interaction Hamiltonian Eq.(1), we next apply DQMC to unveil its physical behavior quantitatively. To characterize the AF order, we calculate the spin structure factor, which is defined by~\cite{Varney2009},
\begin{align}\label{eq1}
S(\mathbf{q},L)=\frac{1}{N_s^2}\sum_{i,j}e^{i\mathbf{q}\cdot(\mathbf{r}_i-\mathbf{r}_j)}\langle {\bf S}_i \cdot{\bf S}_j \rangle
\end{align}
where the spin operator is ${\bf S}_i=(S^x_i, S^y_i, S^z_i)$; $N_s$ is the total number of sites. The antiferromagnetism has an order vector ${\bf q}_0=(\pi,\pi)$, and we let $S_{AF}=S({\bf q}_0)$. For the plaquette-dimer phase, we define the following static structure factor,
\begin{eqnarray}\label{pifluxdimer}
S_{DM}(L)=\frac{1}{N^2_s}\sum_{i,j}\sum_{\sigma,\sigma'}\langle \Delta_{i\sigma}\Delta_{j\sigma'}^{\dagger}\rangle,
\end{eqnarray}
where the bond operator writes as
\begin{align}\label{eq1}
\Delta_{i\sigma} & =-[(-1)^{i_x+1}t_{i,i+x}\{(c^{\dagger}_{i\sigma}c_{i+x\sigma}-c^{\dagger}_{i\sigma}c_{i-x\sigma})+\textrm{H.c.}\} \nonumber\\
& +i(-1)^{i_y+1}t_{i,i+y}
\{(c^{\dagger}_{i\sigma}c_{i+y\sigma}-c^{\dagger}_{i\sigma}c_{i-y\sigma})+\textrm{H.c.}\}].
\end{align}

\begin{figure}[htbp]
\centering \includegraphics[width=8.cm]{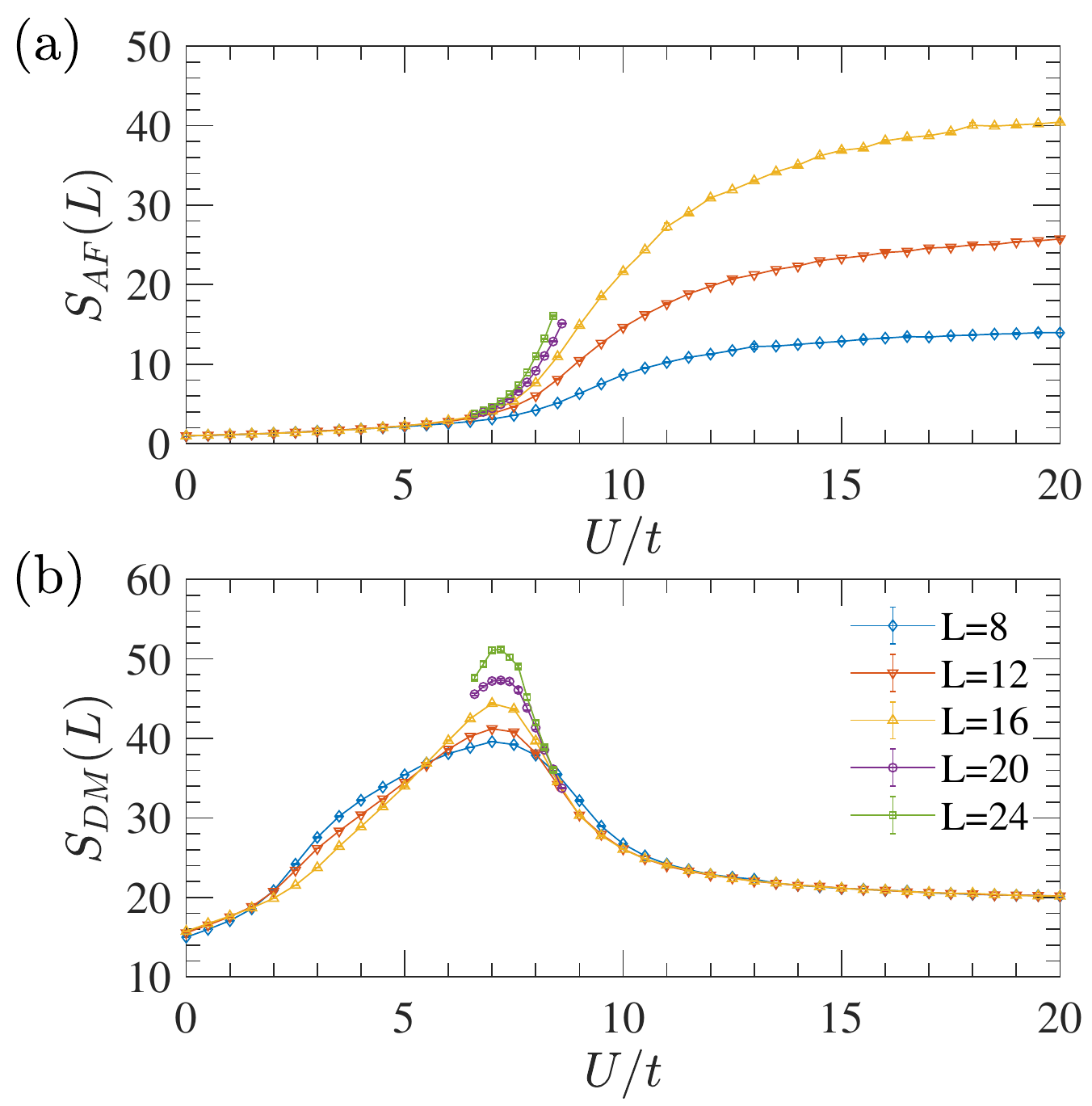} \caption{The structure factors of the two kinds of orders as a function of $U$ for various lattice sizes: (a) the AF structure factors $S_{AF}(L)$; (b) the plaquette-dimer structure factor $S_{DM}(L)$.}
\label{kanspinkekule}
\end{figure}

\begin{figure}[htbp]
\centering \includegraphics[width=9.cm]{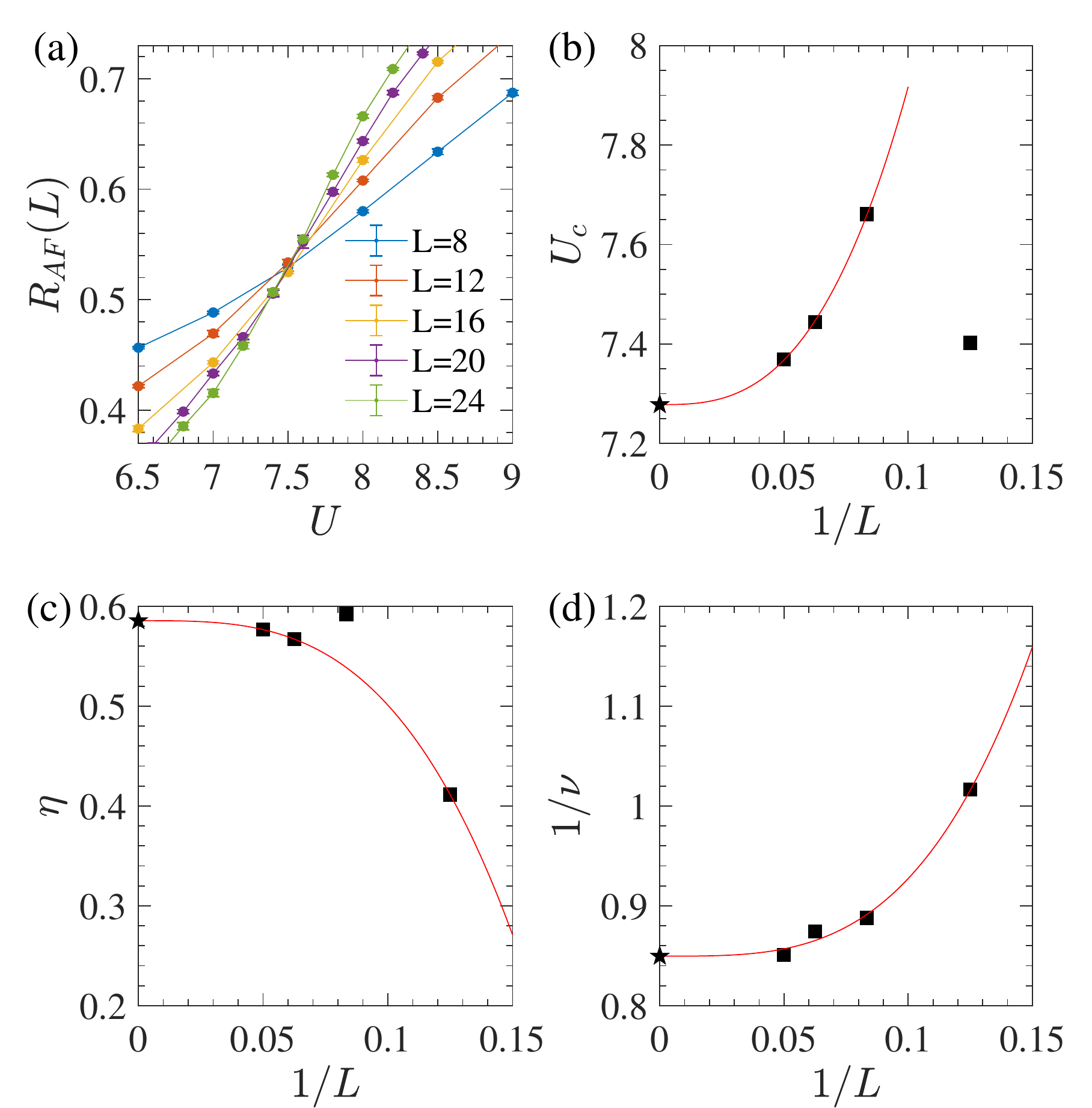} \caption{(a) The RG-invariant correlation ratio of the AF structure factor as a function of the cluster-charge interaction $U$ for various lattice sizes. (b) The critical value $U_c(L)$ determined by the crossing of two $R_{AF}(L)$ curves with consecutive sizes. The anomalous dimension $\eta(L)$ (c) and the correlation function exponent $\nu$ (d) calculated according to Eq.(23). The red solid vurves in (b),(c), and (d) are from polynomial fitting schemes. The fitted values in the thermodynamic limit are $U_c/t=7.278$ ,$\eta = 0.5856$, and $1/\nu = 0.8496$, respectively. }
\label{spinscale}
\end{figure}

In Fig. \ref{kanspinkekule}(a), we plot $S_{AF}(L)$ as a function of the cluster-charge interaction $U$ for various values of $L$. The AF structure factor increases continuously with $U$, and tends to be constant for large enough $U$. Besides, the saturated value increases with $L$. These suggest that the AF order develops at large $U$ and the AF transition is continuous.
To determine the critical interaction strength, we compute the renormalization-group (RG) invariant ratio of the AF structure factor~\cite{Scalettar2019,Meng2019}:
\begin{align}\label{eq1}
R_{AF}(L)=1-\frac{S_{AF}(\mathbf{q}_0+\delta \mathbf{q},L)}{S_{AF}(\mathbf{q}_0,L)},
\end{align}
where $\delta \mathbf{q}$ points to a NN momentum in the Brillouin zone. In the presence (absence) of long-range AF order, we have $S_{\mathrm{AF}}(\mathbf{q}_0+\delta \mathbf{q}) \rightarrow 0\left[S_{\mathrm{AF}}(\mathbf{q}_0]\right)$, and thus $R_{AF}(L) \rightarrow 1(0)$. At the critical point, the use of $R_{AF}$ is advantageous as it has smaller scaling corrections than $S_{\mathrm{AF}}(\mathbf{q})$ itself.
Moreover $R_{c}$ has no scaling dimension, thus it will cross at the critical point $U_c$ for different system size $L$.
However due to the finite-size effect, the curves of $R_{AF}$ for different lattice sizes do not cross exactly at the same point [see Fig. \ref{spinscale}(a)]. Figure \ref{spinscale}(b) shows the critical value $U_c(L)$ determined by the crossing of two consecutive sizes, i.e., $L$ and $L+4$. By extrapolating to the thermodynamic limit, the critical interaction is estimated to be $U_c/t= 7.278$.

The universal scaling functions describing the AF structure factor and RG-invariant correlation ratio around the quantum critical point are\cite{Li2019}:
\begin{align}\label{SR}
S_{AF}(L) & =L^{-(d+z-2+\eta)}F_1[(U-U_c)/U_c\cdot L^{1/\nu},L^{-b_1}],   \nonumber\\
R_{AF}(L) & =F_2[(U-U_c)/U_c \cdot L^{1/\nu},L^{-b_2}],
\end{align}
where the critical exponent $\eta$ is anomalous dimension, and $\nu$ is correlation function exponent; $d$ is space dimension exponent which $d=2$, and $z$ is dynamical critical exponent which is $z=1$ due to the Lorentz invariant; the terms $L^{-b_1}$ and $L^{-b_2}$ are subleading finite-size correlations; $F_1$ and $F_2$ ara unknown ansatz scaling functions. Based on above scaling function, we can extract the values of $\eta$ and $\nu$:
\begin{align}\label{eq1}
\eta(L) & =\frac{1}{\textrm{log}\left(\frac{L}{L+4}\right)}\left.\textrm{log}\left(\frac{S_{AF}(L+4)}{S_{AF}(L)}\right)\right|_{U=U_c}-(d-1), \nonumber\\
\frac{1}{\nu(L)} & =\frac{1}{\textrm{log}\left(\frac{L+4}{L}\right)}\left.\textrm{log}\left(\frac{\frac{d}{dU}R_c(L+4)}{\frac{d}{dU}R_c(L)}\right)\right|_{U=U_c}.
\end{align}

Figure 5(c) and (d) show $\eta(L)$ and $\frac{1}{\nu(L)}$ as a function of inverse lattice size, respectively. The critical exponents in the thermodynamic limit can then be obtained by fitting the data points. The anomalous dimension is determined to be $\eta=0.5856$, and the correlation function exponent is $1/\nu=0.8496$. As a further check, we demonstrate the data collapses of $S_{AF}(L)$ and $R_{AF}(L)$ in Fig.6 (a) and (b), respectively, which are satisfactory for relatively large lattice sizes.

As shown in Fig.4 (b), the plaquette-dimer phase exists in a narrow region between the Dirac semimetal and AF insulator, thus its effective scalings require larger lattice sizes, which is beyond our present computing capacity.  Nevertheless, we try to collapse $S_{DM}(L)$ and $R_{DM}(L)$ using the critical interaction and exponents obtained from the AF transition. The good collapses of the large-size data [see Fig.6(c) and (d)] suggest the critical exponents, obtained from the correlation functions of the two orders with different symmetries, may be the same. Considering the continuous nature of the phase transition, it is highly expected that a DQCP exists between the plaquette-dimer and AF phases.

\begin{figure}[htbp]
\centering \includegraphics[width=8.cm]{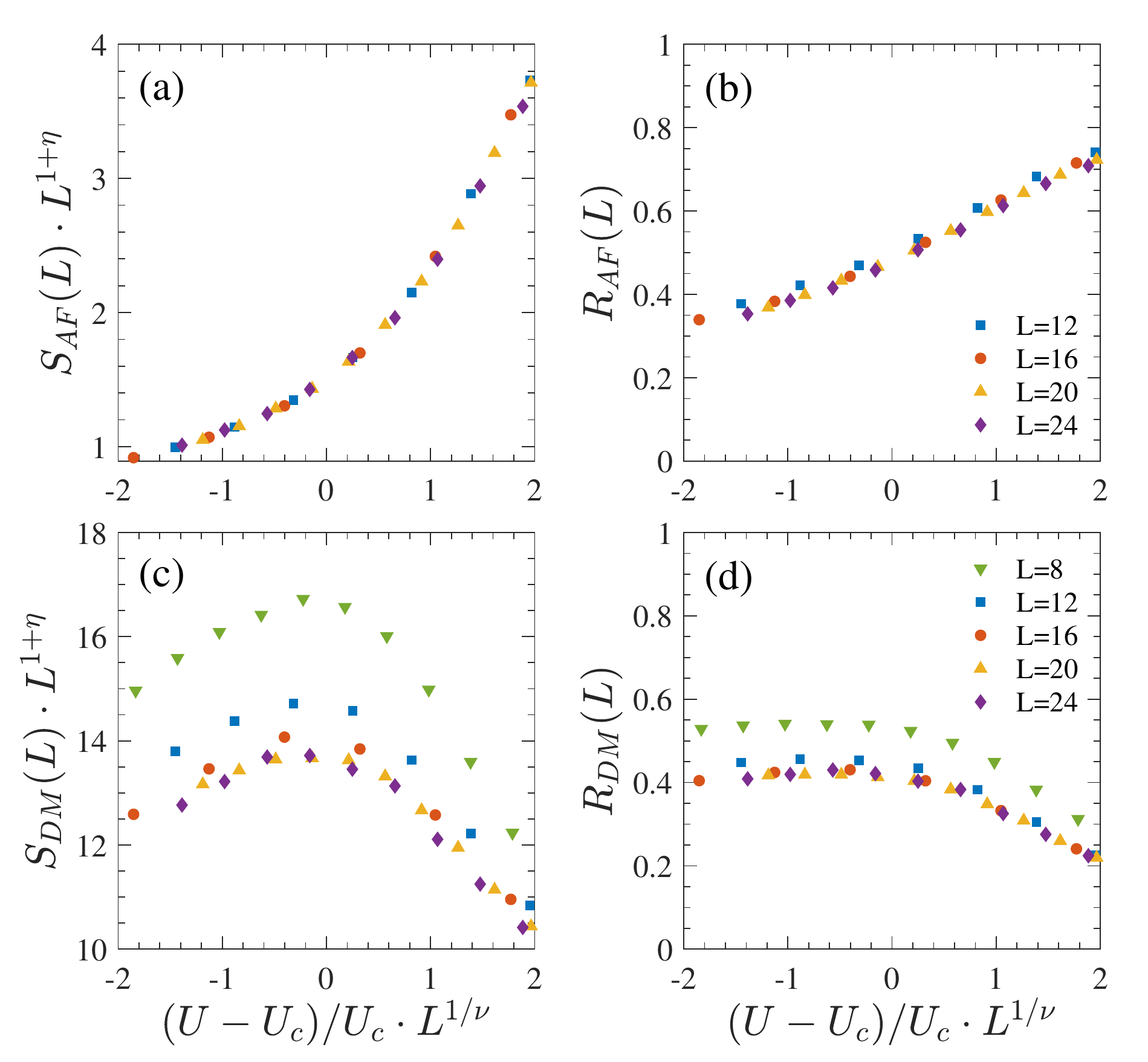} \caption{Data collapses using the critical interaction and exponents determined in Fig.5. (a) The AF structure factor, and (b) the corresponding RG-invariant correlation ratio. (c) and (d) are the plaquette-dimer structure factor and its RG-invariant ratio.}
\label{SRfig}
\end{figure}

\section{Conclusions}
We investigate a specific extended Hubbard model on the $\pi$-flux lattice, in which the interaction terms can be written in a perfect square form of the cluster charge. The mean-field theory predicts a plaquette-dimer phase to occur at a finite interaction strength. While the existence of such a phase is verified by DQMC simulations, it only extends over a narrow parameter region interpolating between the Dirac semimetal and the AF insulator, thus is relatively weak for the lattice sizes accessible by DQMC. In contrast, the AF transition reflected in the spin structure factor is much more obvious, and the critical interaction and exponents are steadily obtained by finte-size scalings. We then find the plaquette-dimer structure factor of large lattice sizes can be well scaled using the above critical values. Our results reveal that a possible DQCP may be induced by the cluster-charge interaction on the $\pi$-flux lattice.
Clearly, DQCP identified here needs further confirmations, either by simulating larger lattice sizes or stabilizing the plaquette-dimer phase with additional interactions, which we leave for future studies.

\textit{Note added.—}While preparing this manuscript, we
noticed a related investigation by Liao et al. \cite{Liao2022} .

\section{Acknowledgments}
H.G. acknowledge support from the National Natural Science Foundation of China (NSFC) grant Nos.~11774019 and 12074022, the NSAF grant in NSFC with grant No. U1930402, the Fundamental Research
Funds for the Central Universities and the HPC resources
at Beihang University.
S.F. is supported by the National Key Research and
Development Program of China under Grant No. 2021YFA1401803,
and NSFC under Grant Nos. 11974051 and 11734002.

\nocite{*}
\bibliography{piflux_ref}

\appendix

\end{document}